\documentstyle[11pt,moriond,epsfig]{article}
\def\Journal#1#2#3#4{{#1} {\bf #2}, #3 (#4)}

\def\NPB{{\em Nucl. Phys.} B}
\def\PLB{{\em Phys. Lett.}  B}
\def\PRL{\em Phys. Rev. Lett.}
\def\PRD{{\em Phys. Rev.} D}
\def\ZPC{{\em Z. Phys.} C}

\def\be{\begin{equation}}
\def\ee{\end{equation}}
\def\bea{\begin{eqnarray}}
\def\eea{\end{eqnarray}}

\begin{document}
\vspace*{4cm}
\title{WW DECAYS AND BOSE-EINSTEIN CORRELATIONS\footnote{talk presented
by K. Fia{\l}kowski at the XXXIIIrd Moriond meeting, "QCD and High Energy 
Hadronic Interactions", Les Arcs, France, March 21-28, 1998. }}

\author{K. FIA{\L}KOWSKI and R. WIT}

\address{Institute of Physics, Jagellonian University,\\
Reymonta 4, 30-059 Krakow, Poland}

\maketitle\abstracts{
Various methods of implementing the Bose-Einstein effect into Monte
Carlo generators, especially for the process $e^+ e^- \rightarrow W^+ W^-$,
are briefly reviewed and their predictions for the W mass shifts are
compared. The weight methods, which yield very similar predictions
independent on the detailed prescription for weights, are discussed
in more detail. In particular, we advocate a new method,
which seems to be practical and reasonably well justified theoretically.}
\pagebreak

\section{Introduction}
With the increasing domination of the models equipped with Monte Carlo
generators, the problem of proper description of the Bose-Einstein
second order interference effect (called often "HBT effect"~\cite{hbt}, and
here denoted as the BE effect)  has reappeared. This effect allows to
learn more about the space-time development of the production processes,
especially if results for semi-inclusive samples of data could be compared
with predictions resulting from various model assumptions.
However, it is non-trivial to implement the quantum interference effect
into Monte Carlo generators, which deal with probabilities and not with
amplitudes.
\par
The problem became suddenly quite acute, when one realized that the
interference effects may result in the mass shifts for W bosons produced
pairwise in $e^+ e^-$ collisions and decaying into hadrons. Conflicting
predictions for this
shift were presented~\cite{web}, resulting in some confusion about the
possibility of using 4-jet events for precise measurements of the W mass.
\par
In this paper we review very shortly different procedures implementing the
BE effect into Monte Carlo generators and compare their
predictions for W mass shifts. We discuss in more detail the
weight methods, presenting a practical algorithm which avoids some of
the difficulties inherent for this class of procedures. We conclude with
the list of further studies to be performed.

\section{How to implement the Bose-Einstein effect in Monte Carlo
generators?}
The standard discussion of the BE effect~\cite{bgj} starts
from the classical space-time source emitting identical bosons with
known momenta. Thus the most natural procedure is to treat the original
 Monte Carlo generator as the model for the source and to symmetrize the
final state wave function~\cite{sul}. This may be done in a more proper way
using the formalism of Wigner functions~\cite{zha}. In any case, however,
the Monte Carlo generator should yield both the momenta of produced
particles and the space-time coordinates of their creation (or last
interaction) points. Even if we avoid troubles with the uncertainty principle
by using the Wigner function approach, such a generator seems reliable
only for heavy ion collisions. It has been constructed also for the $e^+ e^-$
collisions~\cite{eg}, but localizing the hadron creation point in the
parton-based Monte Carlo program for lepton and/or hadron collisions is
a rather arbitrary procedure, and it is hard to say what does one really
test comparing such a model with data.
\par
The best procedure seems
to be taking into account the interference effects before generating
events.  Unfortunately, this was done till now only for the JETSET
generator for a single Lund string~\cite{ar,ar2}, and a generalization
for multi-string processes is not obvious. No similar modifications were
yet proposed for other generators.
\par The most popular approach,
applied since quite a few years to the description of BE effect in various
processes, is to shift the final state momenta of events generated by the
PYTHIA/JETSET generators~\cite{sjo}.  The prescription for a shift starts
from the observation that original generators produce very small
correlations in two-particle distributions of like-sign pairs of pions.
The standard quantity to measure such correlations is the ratio
\begin{equation} c_2(Q)=\frac{<n>^2}{<n(n-1)>} \frac{\int d^3 p_1 d^3 p_2
\rho_2 (p_1,p_2) \delta (Q-\sqrt{-(p_1-p_2)^2})} {\int d^3 p_1 d^3 p_2
\rho_1 (p_1) \rho_1 (p_2) \delta (Q-\sqrt {-(p_1-p_2)^2})},
\label{eq:ratio}
\end{equation}
\noindent
which is a function of a single invariant variable $Q$. As noted above, this
ratio is close to one for a default generator version, whereas experiments
show the "BE enhancement", often parametrized by

\begin{equation}
c_2(Q)=1+\lambda exp(-R^2 Q^2),
\label{eq:factor}
\end{equation}
\noindent
where $R$ and $\lambda$ are parameters interpreted as the source radius and
"incoherence strength", respectively.
\par
Now for all the pairs of identical pions\footnote{In fact, only the
direct pions and $\rho$ decay products are counted, since for other pairs
the effective source size is too large to give the visible enhancement in
momentum space.} a shift of momenta is calculated
to assure such a shift in the value of $Q=\sqrt{-(p_1-p_2)^2}$, that the
resulting numerator of ~(\ref{eq:ratio}) will be multiplied by the
"BE factor" ~(\ref{eq:factor}). The shift of momenta is made unique by
the requirement that the  pair's 3-momentum should not change. After
performing the shifts, all the CM 3-momenta of final state particles are
 rescaled to restore the original energy. In more recent versions of the
procedure~\cite{ls2} "local rescaling" is used instead of the global one.
In any case, each event is modified and the resulting generated sample
exhibits now the "BE enhancement": the ratio ~(\ref{eq:ratio}) is no longer
close to one, and may be parametrized as in ~(\ref{eq:factor}).  \par There
is no theoretical justification for the procedure, so it should be
regarded as an imitation rather than implementation of the BE
effect. Its success or failure in describing data is the only relevant
feature. Unfortunately, whereas the method is very useful for the
description of two-particle inclusive spectra, it fails to reproduce
(with the same fit parameters $R$ and $\lambda$) the three-particle
spectra~\cite{ua1} and the semi-inclusive data~\cite{na22}. This could be
certainly cured, e.g., by modifying the shifting procedure and fitting the
parameters separately for each semi-inclusive sample of data. However, the
fitted  values of parameters needed in the input factor ~(\ref{eq:factor})
used to calculate shifts are quite different from the values one would get
fitting the resulting ratio ~(\ref{eq:ratio}) to the same form~\cite{fw1}.
Thus it seems to be very difficult to learn something reliable on the
space-time structure of the source from the values of fit parameters in
this procedure.
\par All this has led to the revival of weight methods,
known for quite a long time~\cite{pra}, but plagued with many practical
problems. The method is clearly justified within the formalism of the
Wigner functions, which allows to represent (after some simplifying
assumptions) any distribution with the BE effect built in as a product of
the original distribution (without the BE effect) and the weight factor,
depending on the final state momenta~\cite{bk}. With an extra assumption
of factorization in momentum space, we may write the weight factor for
final state with $n$ identical bosons as \begin{equation}
W(p_1,...p_n)=\sum \prod_{i=1}^n w_2(p_i,p_{P(i)}), \label{eq:weight}
\end{equation} \noindent where the sum extends over all permutations
${P_n(i)}$ of $n$ elements, and $w_2(p_i,p_k)$ is a two-particle weight
factor  reflecting the effective source size. A commonly used simple
parametrization of this factor for a Lorentz symmetric source is
\begin{equation}
w_2(p,q)=exp[-(p-q)^2R^2/2],
\label{eq:wf}
\end{equation}
The only free parameter is now $R$, representing the effective source size.
In fact, the full weight given to each event should be a product of
factors ~(\ref{eq:weight}) calculated for all kinds of bosons; in practice,
pions of all signs should be taken into account. As before, only direct
pions and $\rho$ decay products should be taken into account, since for
other pairs much bigger $R$ should be used, resulting in negligible
contributions.
\par
There are two problems with using ~(\ref{eq:weight}) as a prescription
for weight to be given to each generated event. First, as the sum contains
$n!$ terms, the time needed for its calculation becomes prohibitive for
more than 15 identical particles in the final state. This has been dealt
with in different ways: by allowing for permutation only in separate CM
hemispheres~\cite{hay}, by replacing the sum ~(\ref{eq:weight}) by other,
better or worse justified ansatzes~\cite{jz,kkm,tnr},
or by restricting the class of permutations, over which the sum is
performed~\cite{fw2,fw3}. It is rather difficult to judge,
how precise is the approximation of the sum ~(\ref{eq:weight}) in each
case, although for the last methods some estimates were given~\cite{fw2}.
We will return to this problem later on.
\par
The second problem concerns side-effects of the weights. Obviously,
introducing weights changes not only the ratio ~(\ref{eq:ratio}), but
all the distributions obtained from the generated sample of events. In
particular, since the values of weight factors ~(\ref{eq:weight}) will be
in average larger for the larger multiplicities, the multiplicity
distribution may be seriously distorted by weights. Let us stress that
this is by no means a drawback of the weight methods: in the real world
there is always a BE effect, and if the free parameters in Monte Carlo
were fitted to the data neglecting this effect, their values are simply
incorrect. However, the iterative procedure of refitting the parameters
taking the weights into account would be very tedious. Thus a simple
ansatz may be used~\cite{jz}: the calculated weights should be rescaled
by a simple factor $cV^n$, where $n$, as above, is the number of
identical bosons in the final state, and $c$, $V$ are free parameters.
Their values should be fitted to restore the shape and normalization of
the original multiplicity distribution. Obviously, this ansatz may be
insufficient for a more detailed analysis. For example, since in general
different parameters define the distribution of the number of jets $N_j$
 in the $e^+ e^-$ collisions, and the distribution in number of particles
$j$ for a single jet, double rescaling in $N_j$ and $j$ may be needed if
one wants to analyze the fully inclusive sample. The energy dependence
may be also troublesome. Nevertheless, for the particular problem of the $WW$
pair production the present versions of the weight methods were found to
be sufficient~\cite{jz,fw3}.

\section{Predictions for the W mass shifts}
The four classes of the procedures outlined in the previous section give
very different predictions for the $W$ mass shift in the $WW \rightarrow
4 jets$ final state. There are not many really new results since last year,
so we may refer the reader for more detailed analysis to the review paper
by Webber~\cite{web}. Here we give only a very short recapitulation
supplemented by a few new developments.
\par
Very large mass shifts (hundreds of MeV) are predicted for symmetrized
production from the parton cascade~\cite{eg}, but this comes mainly from the
unorthodox color reconnection effects, and not from the BE effect. This model
seems to be already contradicted by the data~\cite{hoo}.
\par
Results for the Lund string with interference has been recently presented.
Perhaps not surprisingly, taking into account the BE effect inside each
string does not result in any significant mass shift~\cite{hr}.
\par
Various weight methods, although differing significantly in the
prescriptions for weights and their spectrum, predict also negligible mass
shifts (below 20 MeV)~\cite{jz,kkm,tnr,fw3}. Let us note that this is not
trivial: the weights could be {\it a priori} correlated with the 2-jet
mass value, resulting in a quite substantial mass shift.
\par
The momentum
shifting method~\cite{ls} predicts mass shift about 200 MeV, which was
attributed to the global momentum rescaling present in this
procedure~\cite{jz}. The new versions of the procedure, using more local
rescaling~\cite{ls2}, give essentially no predictions, as the values
obtained range from 0 to 180 MeV, depending on the details of the
algorithm.  \par Thus one can see that the excitement over the subject has
been significantly reduced. It seems rather unlikely that the BE effect
should damage the possibility of high precision measurement of the W mass
in 4 jet events. On the other hand, obviously none of the methods
implementing the BE effect in the Monte Carlo programs is fully
satisfactory and really well developed and a lot of work is needed to
solve remaining practical problems. In the next section we will shortly
discuss some recent improvements in the weight methods.

\section{New developments for the weight methods}

As already noted, it is difficult to estimate, how well one approximates
formula ~(\ref{eq:weight}), even if different truncations of the sum seem
to give very similar results\cite{fw2}. Thus it was proposed\cite{wos} to
use an integral representation of this sum, borrowed from the field theory,
and to calculate the integral in the saddle point approximation. There is,
however, a condition for the momentum configuration, for which this method
may be applied: each momentum must be close (in the sense of smallness of
$Q^2=-(p_1-p_2)^2$) to at least one another momentum. Since this is in
general not true for the final states in the multiparticle production, one
must divide first the final state into {\bf clusters} fulfilling this
condition.  The weight factor for each kind of identical particles is then
a product of the weight factors for all clusters. It was found\cite{fww}
that for reasonable values of parameters the clusters contain
typically only one or a few particles. Thus in fact the integral
representation\cite{wos} is not needed: exact calculation of the sum
~(\ref{eq:weight}) for clusters of less than five particles, and a simple
truncation of the sum\cite{fw2} for larger clusters provides a good
approximation for the full weight, although it needs much less computer
time than the previous methods\cite{fww}.
\par
\begin{figure}[h]
\centerline{%
\epsfig{figure=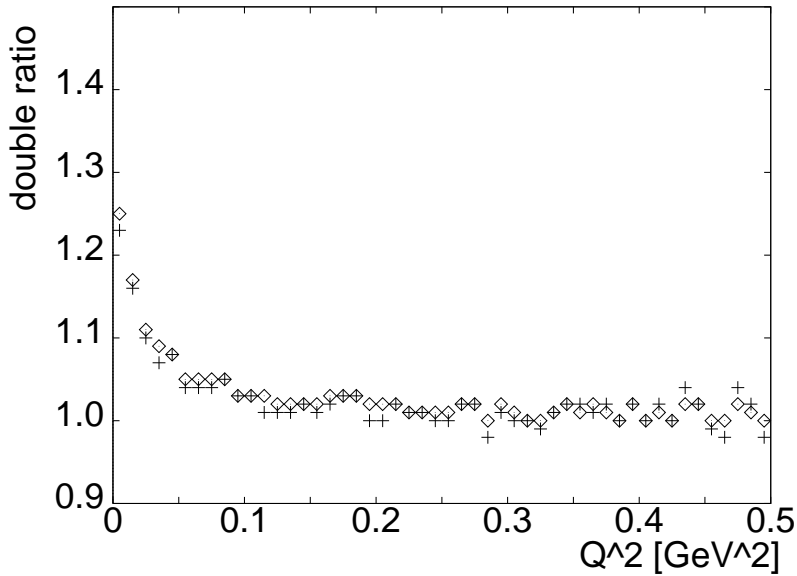,height=7cm}}
\vspace{0.5cm}
\caption{The ratio of "BE ratios" (1) for positive pions with- and without
weights generated for $e^+e^-$ collisons at $Z^0$ peak. Diamonds and
crosses correspond to the rescaled and unrescaled weights, respectively.}
\label{fig:fig1}
\end{figure}

 The weak spot of the
weight method seems to be the rescaling procedure\cite {jz,fw2}. Therefore
it is encouraging to note that the shape of BE ratio ~(\ref{eq:ratio})
seems to depend very weakly on this procedure\cite{fw3} (see
fig.~\ref{fig:fig1}).

 Moreover, we have checked that (at least for the pp collisions) using
the rescaling parameters fitted to restore the original multplicity
distribution one restores as well the original inclusive momentum
distributions\cite{fww}. Thus the simple rescaling seems to work better
than expected.
\par
Let us conclude this section with a remark on the average multiplicities
in the hadronic $W$ decay. Some preliminary data suggest that the multiplicity
for $WW$ final state is more than just twice the multiplicity from a
single W decay\cite{dap}. Such an effect could not be described by the momentum
shifting method\cite{sjo}, since in this case all events preserve their
multiplicities. On the other hand, rescaling of weights which restores the
original distributions for a single $W$ decay will in general change the
average multiplicity for $WW$ final state, since the weight in this case
is not just a product of weights for decay products of two single $W$.
Thus more precise measurement of this effect may decide which method
is better for implementing the BE effect into Monte Carlo generators.
\section{Conclusions and outlook}
We have reviewed shortly the methods of implementing the BE effect into
Monte Carlo generators. One may conclude that the competition for "best
BE in MC" is not yet decided, but may soon be over. There are already
working weight methods which may replace the dominant momentum shifting
method. The future tests should include not only the quantities
relevant for $WW$ production (as the multiplicity shift mentioned above),
but also the problems in which the momentum shifting method has failed,
as the semi-inclusive data and higher order correlations. One should
consider non-symmetric form of two-particle weight factor ~(\ref{eq:factor}),
the dependence of its free parameters on energy and the possibility of
different parameter values for various pairs (e.g. from the same- and from
different $W$-s). In any case, one should stress that we do not speak about
comparing data with non-existing "world without BE effect". There is a
lot of possible real physical effects due to the BE effect (as the difference
between the $WW$ state and the superposition of two single $W$-s) and investigating
them will certainly enlarge our knowledge on the space-time development of
the multiple production processes.

\section*{Acknowledgments}
This work was partially supported by the KBN grants No 2 P03B 086 14 and
No 2P03B 196 09.

\end{document}